\UseRawInputEncoding 
\RequirePackage{fix-cm}
\documentclass[smallcondensed]{svjour3}       % onecolumn (second format)
\smartqed  % flush right qed marks, e.g. at end of proof
\usepackage{graphicx}
\journalname{Computational and Applied Mathematics}
\begin{document}

\title{On retrograde orbits, resonances and stability  %\thanks{Grants or other notes
%about the article that should go on the front page should be
%placed here. General acknowledgments should be placed at the end of the article.}
}
%\subtitle{Do you have a subtitle?\\ If so, write it here}

%\titlerunning{Short form of title}        % if too long for running head

\author{M. H. M. Morais         \and
        F. Namouni %etc.
}

%\authorrunning{Short form of author list} % if too long for running head

\institute{M. H. M. Morais \at
              Instituto de Geoci{\^e}ncias e Ci{\^e}ncias Exatas, Universidade Estadual Paulista (UNESP), Av. 24-A, 1515 13506-900 Rio Claro, SP, Brazil \\
                    \email{helena.morais@rc.unesp.br}           %  \\
%             \emph{Present address:} of F. Author  %  if needed
           \and
           F. Namouni \at
              Universit\'e de Nice, CNRS, Observatoire de la C{\^o}te dÕAzur, CS 34229, 06304 Nice, France
}

\date{Received: 15 April 2015 / Accepted: 1 November 2015}
% The correct dates will be entered by the editor

\maketitle

\begin{abstract}
We  start by reviewing our previous work on retrograde orbital configurations and on  modeling and
identifying retrograde resonances.  Then, we present new results regarding  the enhanced stability of retrograde
configurations with respect to prograde configurations in the low mass ratio regime of the planar circular restricted 3-body problem. Motivated by the recent discovery of small bodies which are in  retrograde resonance with the Solar System's  giant planets we then explore the case with mass ratio $0.001$  and show new stability maps  in a grid of semi-major axis versus eccentricity for the 2/1 and 1/2 retrograde resonances. Finally, we explain how the stability borders of the 2/1 and 1/2 retrograde resonances are related to the resonant orbits' geometry.
% \PACS{PACS code1 \and PACS code2 \and more}
% \subclass{MSC code1 \and MSC code2 \and more}
\end{abstract}

\section{Introduction}
A retrograde orbital configuration consists of two or more bodies moving around a central mass in opposite directions or, more precisely, such that the relative inclination\footnote{The relative inclination between two orbits is the angle between the respective  angular momentum vectors.}  $180^\circ \geq I>90^\circ$. When $I=180^\circ$ the orbits are retrograde and coplanar.  

Examples of retrograde orbital motion in the Solar System include: many comets, 63 small bodies\footnote{Listed on http://www.minorplanetcenter.net/ on 10/4/2015.} including Mars crossing asteroids, a subset of Centaurs called Damocloids, a TNO (2008KV42), and distant objects on nearly parabolic orbits. Many irregular satellites of the giant planets and Neptune's moon  Triton also have retrograde orbital motion.
Extrasolar planetary systems are detected through indirect methods  and the relative inclinations are usually not known\footnote{The radial velocity method usually does not constrain relative inclinations while the transit method only allows to detect nearly-aligned systems.}.  Retrograde configurations could be achieved e.g.\  through capture of a free floating planet in a star cluster which has isotropic probability \cite{Perets2012ApJ}, \cite{Varvoglis2012CeMDA}. 

Analysis of radial velocity data for the star $\nu$-Octantis~A by \cite{Ramm_etal2009MNRAS} led to the claim that there is a planet about half-way between $\nu$-Octantis~A and its binary companion, $\nu$-Octantis~B. This result was puzzling since a planet at such location in a  prograde configuration would be unstable due to the strong perturbation from $\nu$-Octantis~B and led to alternative hypothesis for this system \cite{Morais&Correia2012MNRAS}. Numerical integrations by \cite{Eberle&Cuntz2010ApJ} and \cite{Gozdziewski2013MNRAS} showed that a planet in a retrograde configuration could be stable in the $\nu$-Octantis system which prompted a study of the  stability of retrograde configurations in binary systems by \cite{Morais&Giuppone2012MNRAS}. This showed that the enhanced stability of retrograde configurations with respect to prograde configurations is due to the nature of retrograde resonances which are weaker than prograde resonances at a given orbital period ratio \cite{Morais&Giuppone2012MNRAS}.

In \cite{Morais&Namouni2013CMDA} we studied retrograde resonance in the framework of the circular restricted 3-body problem (CR3BP), identifying the relevant resonant arguments in the disturbing function and exploring retrograde resonance stability using the method of surfaces of section. In \cite{Morais&Namouni2013MNRASL} we identified small bodies (Damocloids) which are currently in the 1/2 and 2/5 retrograde resonances with Jupiter, and in the 2/3 retrograde resonance with Saturn. In  \cite{Namouni&Morais2015MNRAS} we performed a numerical study of the  resonance capture probabilities  for objects on prograde and retrograde orbits migrating inwards towards a Jupiter mass planet. We saw that capture in resonance is more likely for retrograde inclinations ($I>90^\circ$) rather than prograde inclinations  ($I<90^\circ$).

This article is organized as follows: In Sect.~2 we review our work on the retrograde disturbing function, explain how to obtain retrograde resonance's critical angles and associated resonance strength. In Sect.~3 we present  results on the enhanced stability of retrograde configurations with respect to prograde configurations for mass ratio $<0.01$ relevant for planetary systems. In Sect.~4 we show stability maps in a grid of semi-major axis versus eccentricity  for the 2/1 and 1/2 retrograde resonances. In Sect.~5 we discuss our results.  

\section{Retrograde resonance}

\subsection{Retrograde and prograde disturbing functions}

Consider the restricted 3-body problem consisting of a test particle subject to the gravitational influence of a primary and secondary (mass ratio $\mu$) which are on circular orbit of unit radius (CR3BP). 
The  disturbing function is\footnote{We use standard Keplerian elements for the test particle's orbit: $a$ (semi-major axis), $e$ (eccentricity), $I$ (inclination), $f$ (true anomaly), $\omega$ (argument of pericentre), $\Omega$ (longitude of pericentre), $\lambda$ (mean longitude). The secondary's  circular orbit has semi-major axis $a^\prime=1$ and mean longitude $\lambda^\prime$.}
\begin{equation}
R=\mu \left( \frac{1}{\Delta} -r \cos{\psi} \right)
\end{equation}
where $r$ is the radial distance to the primary, $\Delta=\sqrt{1+r^2-2\,r\,\cos{\psi}}$,
\begin{eqnarray}
\cos\psi & = &  \cos(f+\omega) \cos(\Omega-\lambda^\prime) -  \sin(f+\omega) \sin(\Omega-\lambda^\prime) \cos{I} \nonumber \\  
&=& \cos^2(I/2) \cos(f+\omega+\Omega-\lambda^\prime) +   \sin^2(I/2) \cos(f+\omega-\Omega+\lambda^\prime) \ .
\end{eqnarray}
If the orbit is prograde with inclination $I\approx0$ then $\sin^2(I/2)\ll1$ hence we define a small parameter
\begin{equation}
\Psi = \cos{\psi}-\cos(f+\omega+\Omega-\lambda^\prime)     = 2\,  \sin^2(I/2) \sin(\Omega-\lambda^\prime)\sin(\omega+f)  \ ,\label{PSI}
\end{equation}
and expanding the indirect term of Eq.~(1)  around $\Psi=0$ leads to the classical expression \cite{ssdbook}
\begin{eqnarray}
\label{dfunctionp}
\frac{1}{\Delta} &=& \sum_{i=0}^{\infty} \frac{(2 i) !}{(i !)^2} \left( \frac{1}{2} r \Psi \right)^{i}  \frac{1}{\Delta_{0}^{2\,i+1}}
\end{eqnarray}
with $\Delta_{0}=\sqrt{1+r^2-2\,r\,\cos(f+\omega+\Omega-\lambda^\prime)}$.

If the orbit is retrograde with  inclination $I\approx180^\circ$ then $\cos^2(I/2)\ll1$ hence $\Psi\gg 1$ and the expansion in Eq.~\ref{dfunctionp} is not valid. Following \cite{Morais&Namouni2013CMDA} we define a small parameter
\begin{equation}
\bar{\Psi} = \cos{\psi}-\cos(f+\omega-\Omega+\lambda^\prime)     = 2\,  \cos^2(I/2) \sin(-\Omega+\lambda^\prime)\sin(\omega+f) \label{PSI*} \ ,
\end{equation}
and expanding the  indirect term of Eq.~(1)  around $\bar{\Psi}=0$ leads to
\begin{eqnarray}
\label{dfunctionr}
\frac{1}{\Delta} &=& \sum_{i=0}^{\infty} \frac{(2 i) !}{(i !)^2} \left( \frac{1}{2} r \bar{\Psi} \right)^{i}  \frac{1}{\bar{\Delta}_{0}^{2\,i+1}}
\end{eqnarray} 
with $\bar{\Delta}_{0}=\sqrt{1+r^2-2\,r\,\cos(f+\omega-\Omega+\lambda^\prime)}$.

Therefore, we see that the expansions of the disturbing function for prograde (Eq.~\ref{dfunctionp}) and retrograde (Eq.~\ref{dfunctionr}) orbits are related. The latter may be obtained from the former by performing the following canonical transformation \cite{Morais&Namouni2013CMDA}
\begin{equation}
\parbox{2cm}{$I^\star=180^\circ-I$}\quad\quad 
\parbox{2cm}{$\lambda^{\prime\star}=-\lambda^\prime$}\quad 
\parbox{2cm}{$\omega^\star=\omega-\pi$}\quad 
\parbox{2cm}{$\Omega^\star =-\Omega-\pi$} \ .
\label{canonical}
\end{equation}
The canonical transformation (Eq.~\ref{canonical}) has a physical meaning: it is analogous to obtaining a retrograde orbit of inclination $I$ from a prograde orbit of inclination $180^\circ-I$ by simply inverting the motion of the massive bodies.
This is the 3D version of the procedure described in \cite{Morais&Giuppone2012MNRAS} to obtain the retrograde disturbing function from the prograde disturbing function in the planar (2D) problem.

\subsection{Retrograde resonance critical angles}

By inspection of the disturbing function's expansion (Eq.~\ref{dfunctionr}) we conclude that at the $p/q$ retrograde resonance the slow terms are\footnote{The order of a retrograde $p/q$ resonance is equal to  $p+q$ which is the factor of the $\varpi^{\star}$ term in the resonant argument with $k=0$ of the retrograde disturbing function   
\cite{Morais&Namouni2013CMDA}. Conversely, the order of a prograde  $p/q$ resonance is equal to  $p-q$ which is  the factor of the   $\varpi$ term  in the resonant argument with $k=0$ of the prograde disturbing function  \cite{ssdbook}.}
\begin{equation}
e^{p+q-2\,k} \cos^{2\,k}(I/2) \cos(q\,\lambda^{\star}+p\,\lambda^{\prime\star}-(p+q-2\,k) \varpi^{\star}-2\,k\,\Omega^{\star})
\label{resdf}
\end{equation}
with $2\,k\le p+q$, $\lambda^{\star}=\lambda-2\,\Omega$, $\varpi^{\star}=\omega-\Omega=\varpi-2\,\Omega$.

As an example we discuss the case of the 2/1 and 1/2 resonances.  The 2/1 (1/2) prograde resonance has critical angle  $\lambda-2\,\lambda^{\prime}+\varpi$  ($2\,\lambda-\lambda^{\prime}-\varpi$), which appears at 1st order in eccentricity i.e.\ the amplitude (or strength) of the resonance term is 
${\cal O}(e)$, while  the 2/1 and 1/2 retrograde resonances have 2 critical angles each which appear at 3rd order in eccentricity and inclination.
The 2/1 retrograde resonance has critical angles:
\begin{itemize}
\item $\phi^{\star}_{21}=\lambda^{\star}+2\,\lambda^{\prime\star}-3\,\varpi^{\star}=\lambda-2\,\lambda^{\prime}-3\,\varpi+4\,\Omega$,  \quad\quad\quad\quad  ${\cal O}(e^3)$;
\item $\phi^{\star}_{21}=\lambda^{\star}+2\,\lambda^{\prime\star}-\varpi^{\star}-2\,\Omega^{\star}=\lambda-2\,\lambda^{\prime}-\varpi+2\,\Omega$,  \quad  ${\cal O}(e \cos^2(I/2))$;  
\end{itemize}
while the 1/2 retrograde resonance has critical angles:
\begin{itemize}
\item  $\phi^{\star}_{12}=2\,\lambda^{\star}+\lambda^{\prime\star}-3\,\varpi^{\star}=2\,\lambda-\lambda^{\prime}-3\,\varpi+2\,\Omega$   \quad\quad\quad\quad ${\cal O}(e^3)$, ;  
\item $\phi^{\star}_{12}=2\,\lambda^{\star}+\lambda^{\prime\star}-\varpi^{\star}-2\,\Omega^{\star}=2\,\lambda-\lambda^{\prime}-\varpi$, \quad\quad\quad  ${\cal O}(e \cos^2(I/2))$.  
\end{itemize}

\begin{figure*}
\centering
    \includegraphics*[width=6.2cm]{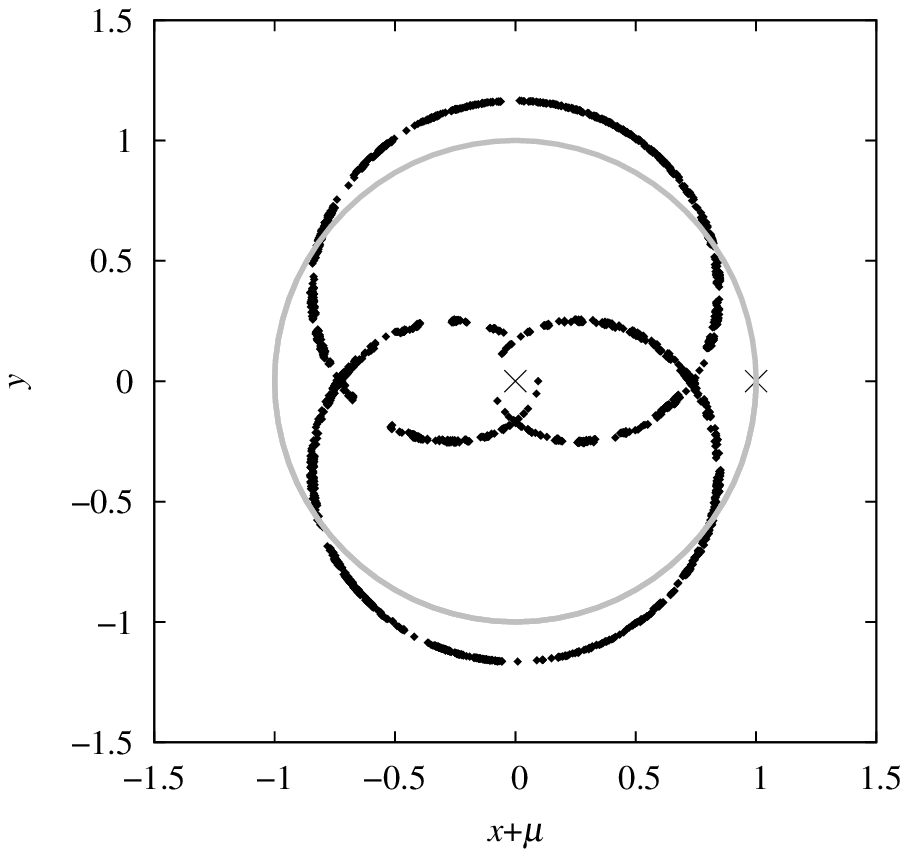}\includegraphics*[width=6.2cm]{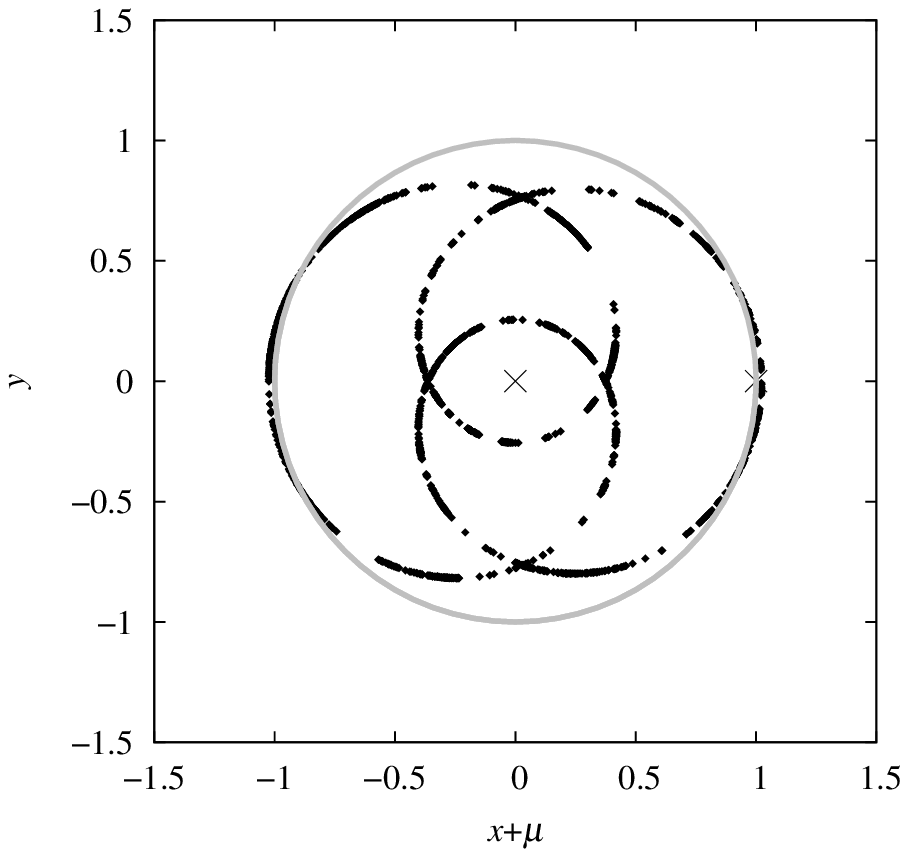}    \\
     \includegraphics*[width=6.2cm]{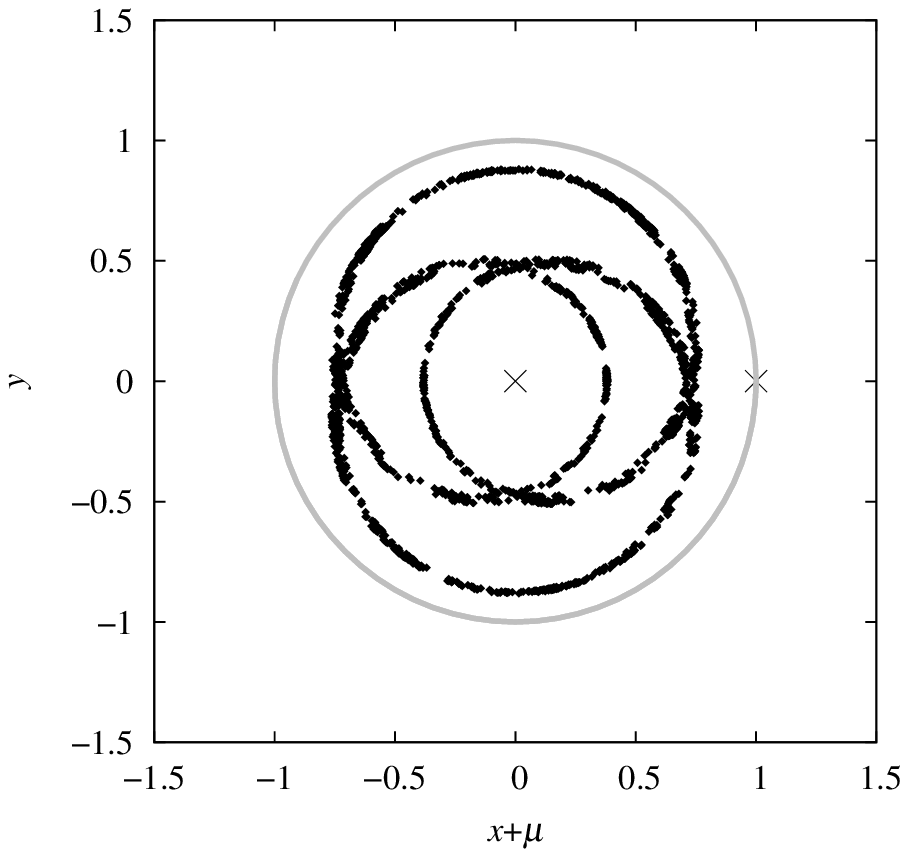}\includegraphics*[width=6.2cm]{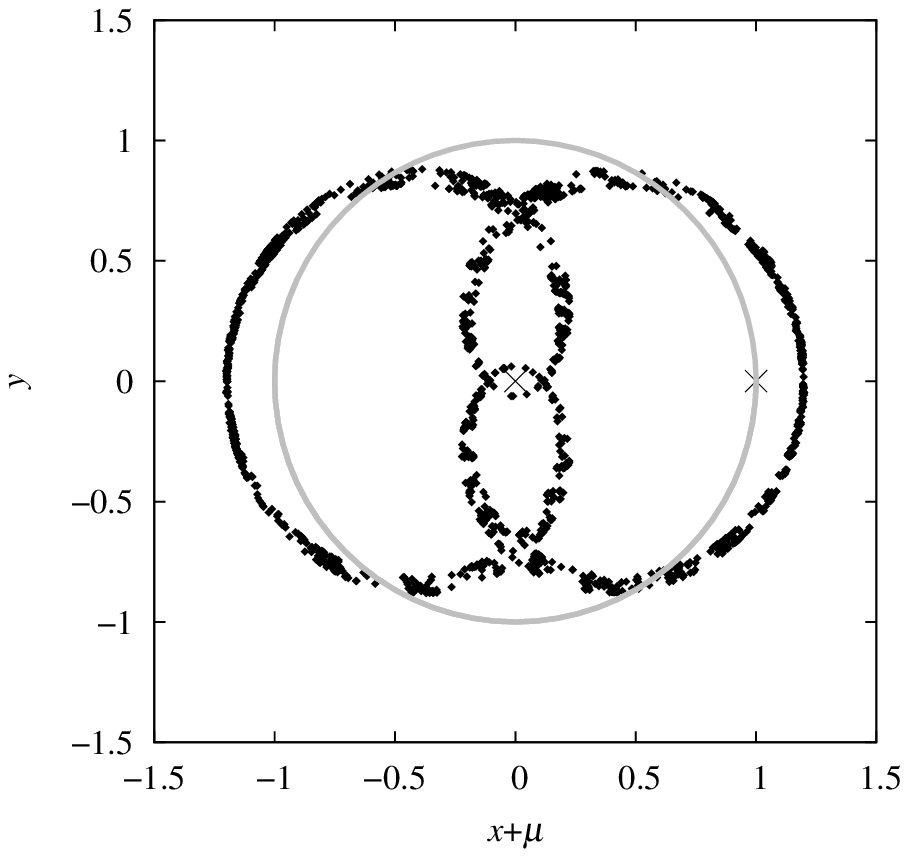} 
\caption{Orbits in 2/1 retrograde resonance (CR3BP with mass ratio $\mu=0.001$): libration around 0 (left) with $a=0.63$ and $e=0.8$ (top) or $e=0.4$ (low);
libration around $180^\circ$ (right) with $a=0.64$ and $e=0.6$ (top) or $a=0.63$ and $e=0.9$ (low). 
The star and planet are identified by symbol $\times$ at $(0,0)$ and $(1,0)$ and the gray circle has unit radius.}    
\label{xy21}
\end{figure*}

\begin{figure*}
\centering
    \includegraphics*[width=6.2cm]{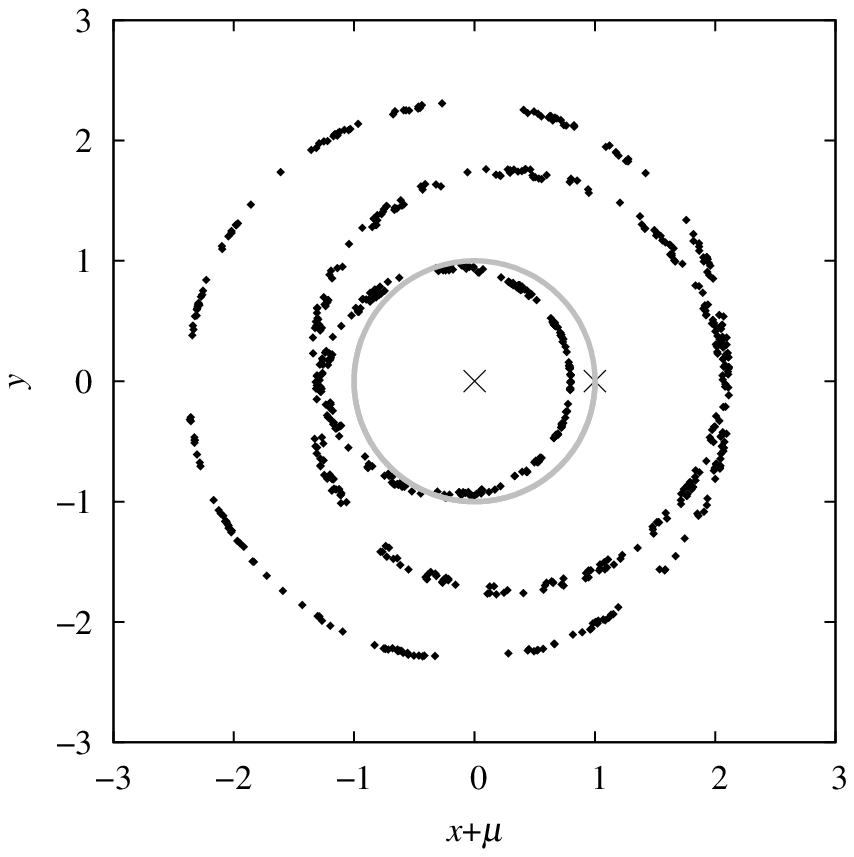}\includegraphics*[width=6.2cm]{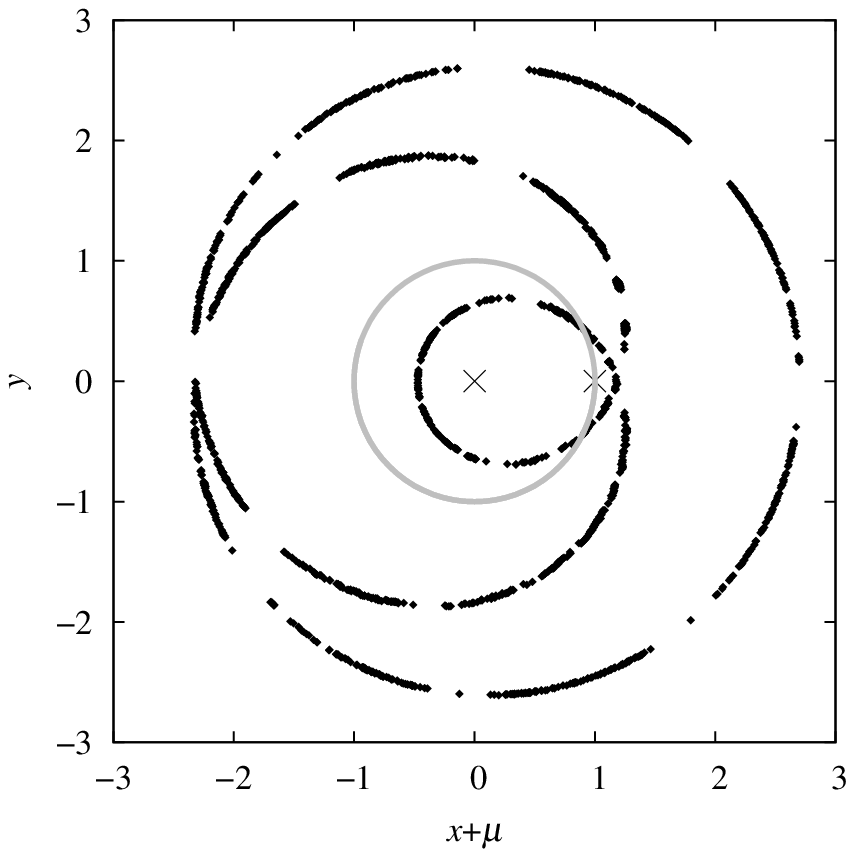}
    \\
      \includegraphics*[width=6.2cm]{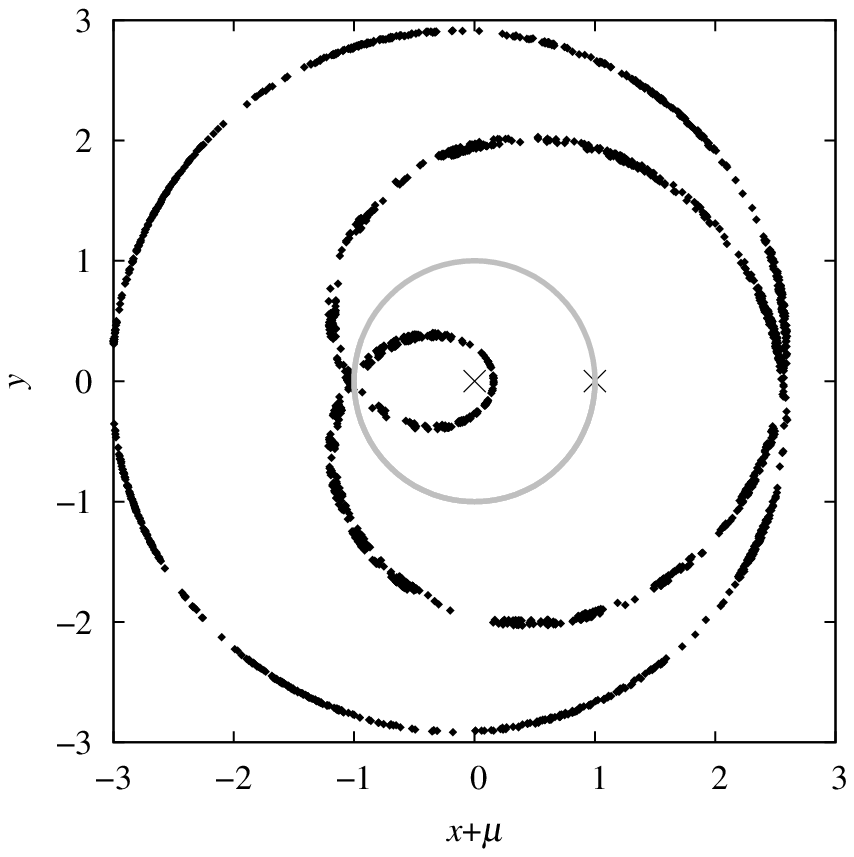}\includegraphics*[width=6.2cm]{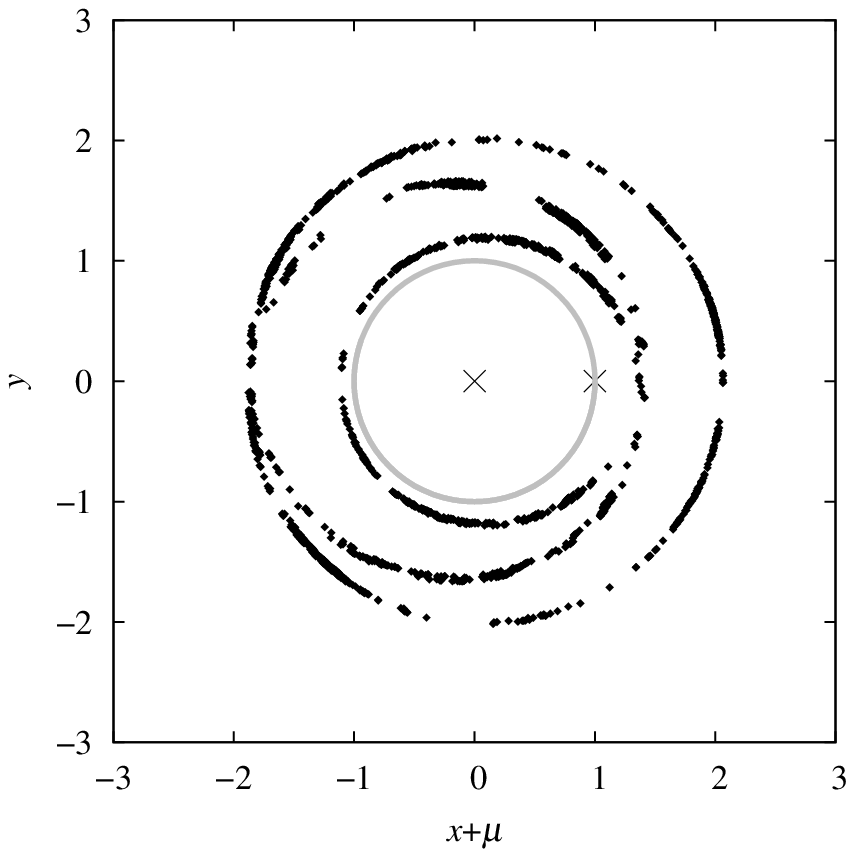}
\caption{Orbits in 1/2 retrograde resonance (CR3BP with mass ratio $\mu=0.001$): libration around 0 (left) with $a=1.6$ and $e=0.5$ (top) or $a=1.59$ and $e=0.9$ (low);
libration around $180^\circ$ (right) with $a=1.59$ and $e=0.7$ (top) or $e=0.3$ (low).
The star and planet are identified by symbol $\times$ at $(0,0)$ and $(1,0)$ and the gray circle has unit radius. }    
\label{xy12}
\end{figure*}

In the planar (2D) problem the 2/1 and 1/2 retrograde resonances have  single critical angles, $\phi^{\star}_{21}=\lambda^{\star}-2\,\lambda^{\prime}-3\,\varpi^{\star}$  and $\phi^{\star}_{12}=2\,\lambda^{\star}-\lambda^{\prime}-3\,\varpi^{\star}$, respectively.
Fig.~\ref{xy21} shows planar orbits in the 2/1 retrograde resonance corresponding to libration of $\phi^{\star}_{21}$ around 0 or $180^\circ$.
Fig.~\ref{xy12} shows planar orbits in the 1/2 retrograde resonance corresponding to libration of $\phi^{\star}_{12}$ around 0 or $180^\circ$.
The location of the resonances' centers at a given mass ratio $\mu$ depends on $a$ and $e$ as we will confirm in Sect.~4.

\section{Enhanced stability of retrograde configurations}

In \cite{Morais&Giuppone2012MNRAS} we showed that a planet orbiting the primary star of a binary may be stable closer to the secondary star if it has a retrograde orbit rather than a prograde orbit. We explained that this is due to retrograde resonances being weaker than prograde resonances, i.e.\ at a given mean motion ratio $p/q$, prograde resonances are of order $p-q$ while retrograde resonances are of order $p+q$. This has  two effects on stability: 1) the excitation of eccentricity is smaller for retrograde resonances; 2) the overlap of resonances which generates a chaotic layer  in the vicinity of the secondary \cite{Wisdom1980AJ} is less widespread for retrograde configurations. This can also be understood in terms of the effect of close encounters with the secondary which generate the chaotic layer in its vicinity. Close encounters in retrograde configurations occur at larger relative velocity (hence have shorter duration and consequentely are less disruptive to the orbit) than in prograde configurations.

In \cite{Morais&Giuppone2012MNRAS} we were interested in the high mass ratio regime and inner orbits, relevant for studying the stability of a planet around the primary star of a binary system. Here, we revisit the problem of stability concentrating in the low mass ratio regime ($\mu<0.01$) and both inner and outer orbits. This is relevant for a small body (test particle)  orbiting a  star subject to the perturbation by a  nearby planet  (mass ratio $\mu$). We numerically integrate the equations of motion of the CR3BP, together with the variational equations and MEGNO\footnote{The fast chaos indicator MEGNO is an acronym for Mean Exponential Growth factor of Nearby Orbits.} equations \cite{Cincotta2000,Gozdziewski2003A&A} for $5\,\times 10^4$ binary's periods.

\begin{figure*}
\centering
  \includegraphics[width=0.85\textwidth]{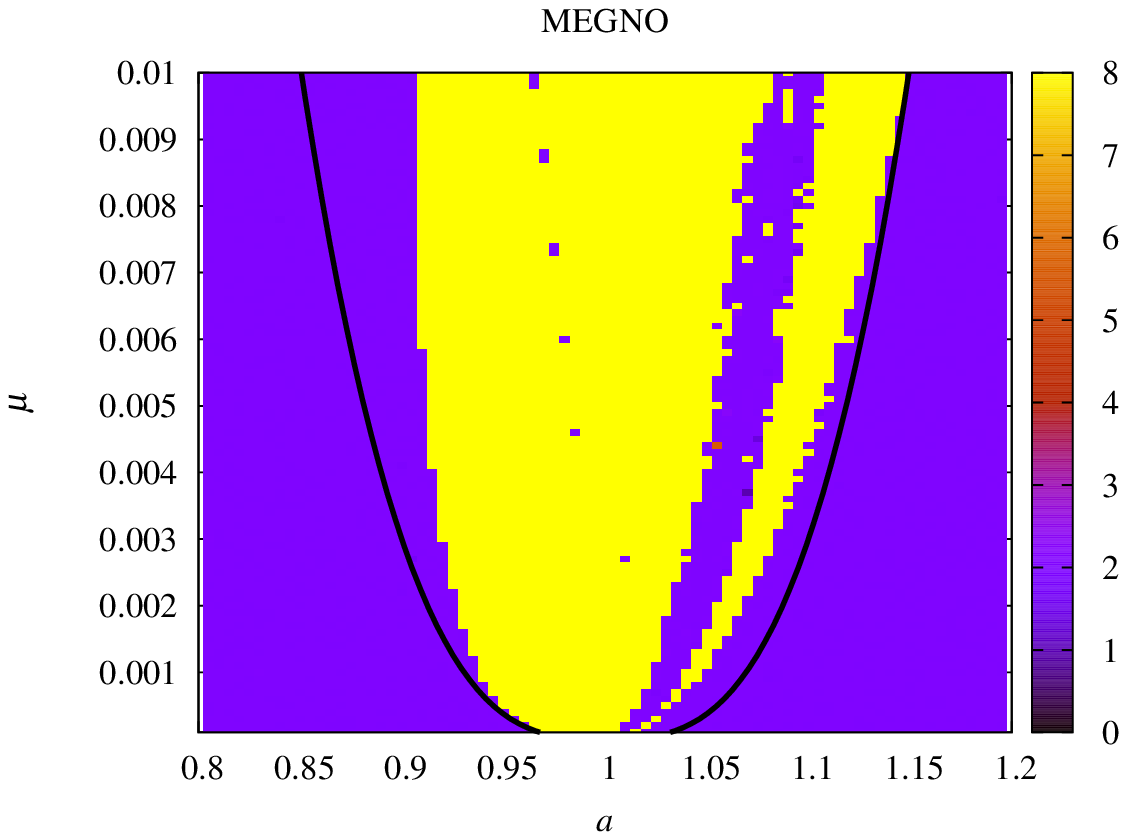} 
  \includegraphics[width=0.85\textwidth]{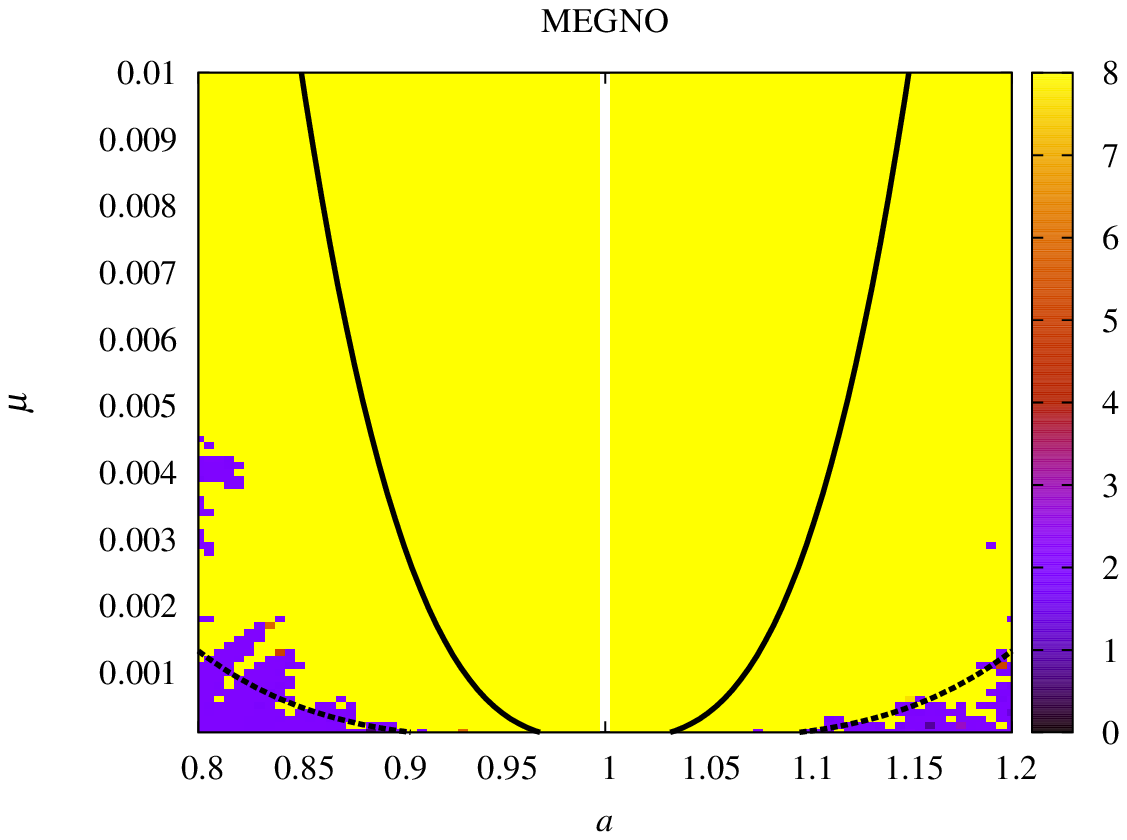}
% figure caption is below the figure
\caption{Stability boundary between regular (blue) and chaotic (yellow) regions for initially circular orbits: retrograde configuration (top panel) and prograde configuration (low panel). Black solid lines show edge of Hill's region (width $(\mu/3)^{1/3}$) while black dashed lines show Wisdom's stability boundary (width $1.33\,\mu^{2/7}$). }
\label{megno}       % Give a unique label
\end{figure*}

In Fig.~\ref{megno} we show the MEGNO maps for retrograde (top panel) and prograde (low panel) configurations in a grid of  semi-major axis  $1.2\geq a \geq 0.8$ varying at steps $0.05\,R_H$,  and mass ratio  $0.01\geq\mu\geq 0.0001$ varying at steps $0.0001$. The other initial orbital elements with respect to the primary were:  $e=0$, $\lambda=\lambda^\prime$, $\omega=0$, $I=180^\circ$ (retrograde configuration) or $I=0$ (prograde configuration). The mean MEGNO $<Y>$ converges to $2$ for regular orbits and increases at a rate inversely proportional to  Lyapunov's time for chaotic orbits  \cite{Cincotta2000,Gozdziewski2003A&A}. We set the maximum $<Y>$ value for chaotic orbits as $8$  in order to have a high contrast between the regular (blue) and chaotic (yellow) regions in Fig.~\ref{megno}.

\begin{figure*}
\centering
  \includegraphics[width=0.95\textwidth]{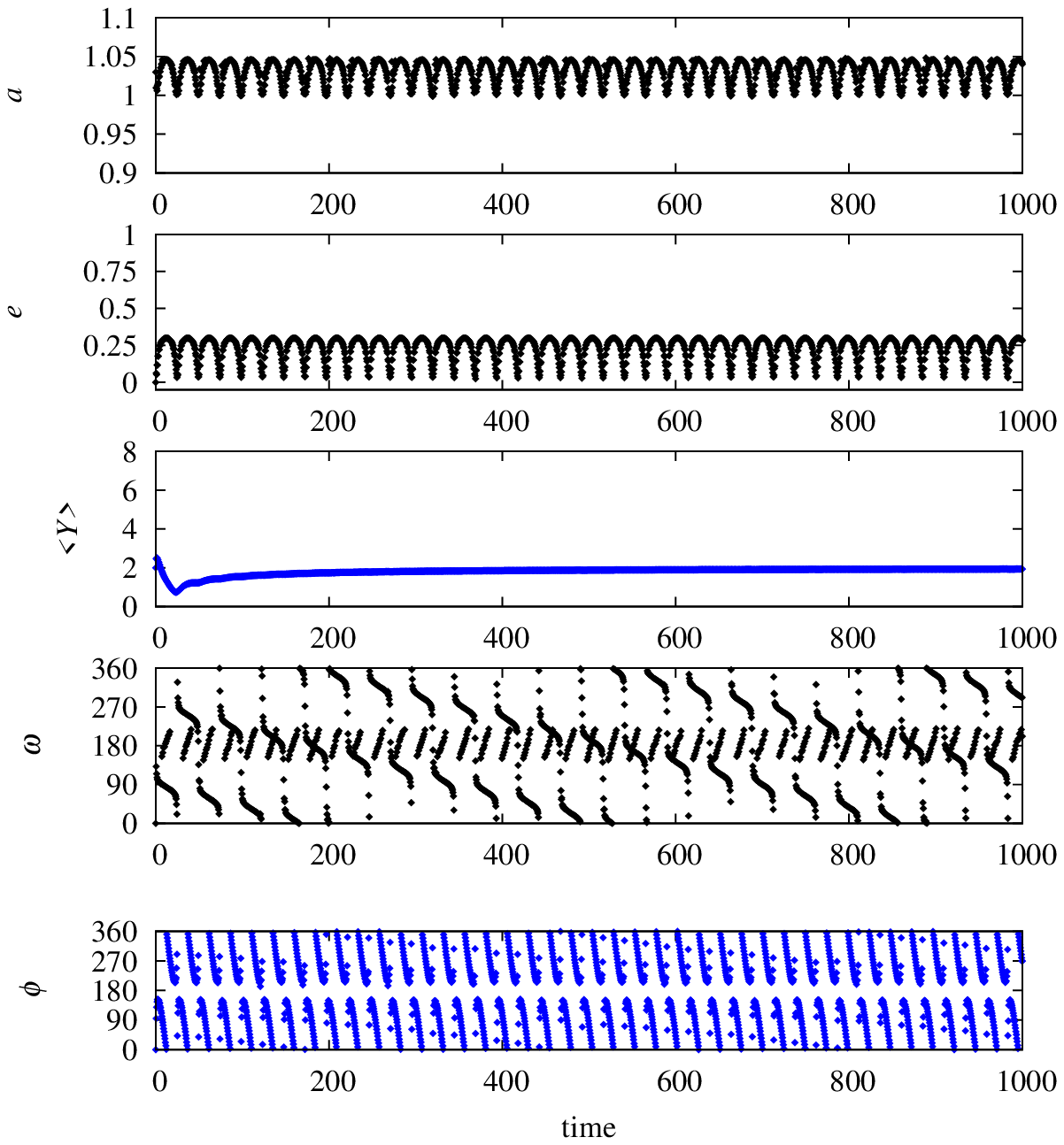}
% figure caption is below the figure
\caption{Orbit within stability island of Fig.~\ref{megno} (top panel) for mass ratio $\mu=0.001$. From top to bottom: semi-major axis $a$, eccentricity $e$, mean MEGNO $<Y>$, argument of pericenter $\omega$, critical angle of 2D retrograde 1/1 resonance $\phi=\lambda-\lambda^\prime-2\,\omega$.}
\label{stabisland}       % Give a unique label
\end{figure*}

Black dashed lines in Fig.~\ref{megno} (low panel) approximate the region of 1st order resonance's overlap for low mass ratio (width $1.33\,\mu^{2/7}$) known as Wisdom's stability criterion \cite{Wisdom1980AJ}. In Fig.~\ref{megno} (low panel) we show only a zoom of the chaotic region with $0.8\le a\le 1.2$ in order to compare with  Fig.~\ref{megno} (top panel).  The boundary of Hill's region (width $R_{H}=(\mu/3)^{1/3}$) is marked by black solid lines in Fig.~\ref{megno}. We see that for prograde configurations, a thick chaotic layer always separates Hill's region from the inner and outer regular regions. This implies that smooth migration capture in the 1/1 prograde resonance is not possible for initial circular orbits outside Hill's region in agreement with our study of resonance capture \cite{Namouni&Morais2015MNRAS}.
In contrast,  for retrograde configurations, Hill's region connects to the inner and outer regular regions. Moreover, in  Fig.~\ref{megno} (top panel) there is striking asymmetry between the inner and outer stability boundaries, both contained within Hill's region. Indeed, the outer boundary has a stability island separating the internal chaotic region from an external chaotic strip. The orbits within the stability island are in the 1/1 retrograde resonance where  the critical angle\footnote{As seen in  \cite{Morais&Namouni2013CMDA} the critical angle of the 2D retrograde 1/1 resonance is $\lambda-\lambda^\prime-2\,\omega$.} $\phi_{11}^{\star}=\lambda-\lambda^\prime-2\,\omega$ librates around 0  with large amplitude (Fig.~\ref{stabisland}). 
 
% For two-column wide figures use

\section{Stability maps of 2/1 and 1/2 retrograde resonances}

As seen in Sect.~2.2 and \cite{Morais&Namouni2013CMDA} the strongest retrograde resonance is the 1/1 (order 2), followed by the 2/1 and 1/2 (order 3).  Here, we explore the stability of the 2/1 and 1/2 retrograde resonances, leaving a detailed study of the 1/1 retrograde resonance for a future publication. To that purpose we fix the mass ratio at $\mu=0.001$ (which represents a small body in the Sun-Jupiter system) and numerically integrate the equations of motion of the CR3BP, together with the variational equations and MEGNO equations  \cite{Cincotta2000,Gozdziewski2003A&A} for $5\,\times 10^4$ binary's periods. As before, we set the maximum mean MEGNO $<Y>$ value for chaotic orbits as $8$  in order to have a high contrast between the regular (blue) and chaotic (yellow) regions.

\subsection{The 2/1 retrograde resonance}

\begin{figure*}
\centering
  \includegraphics[width=0.95\textwidth]{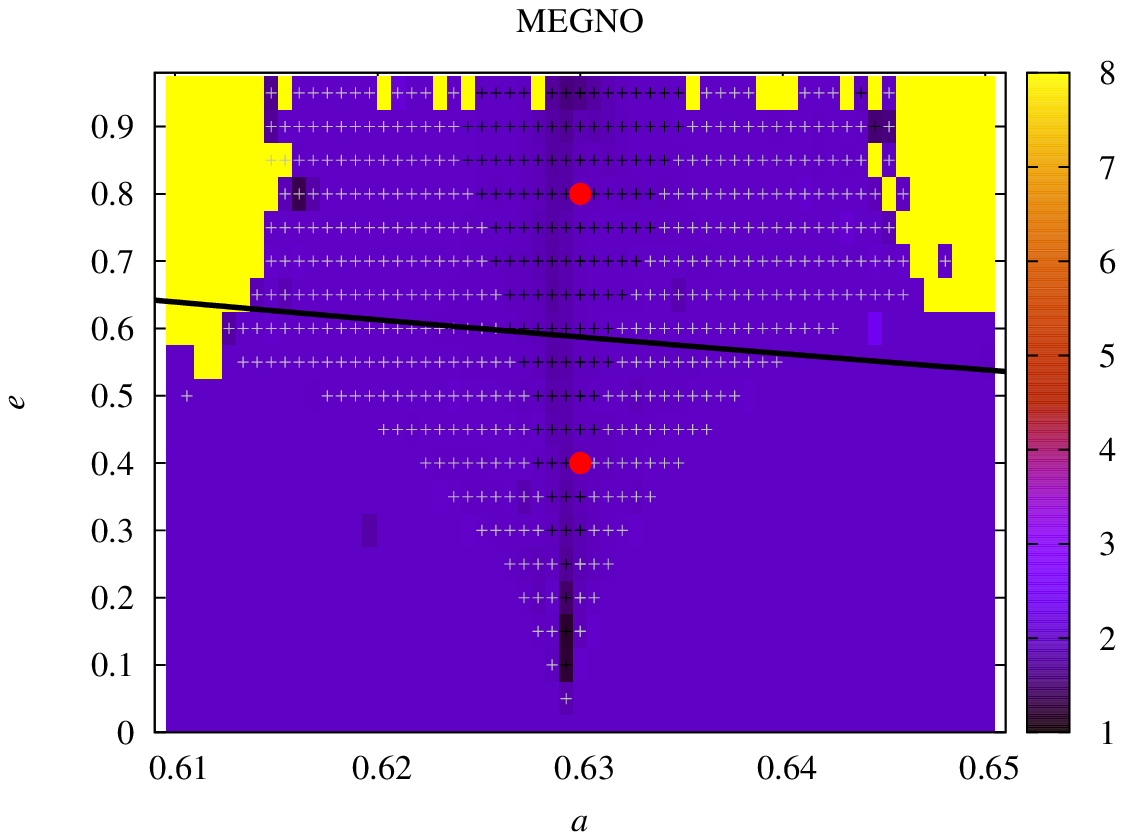}
  \includegraphics[width=0.95\textwidth]{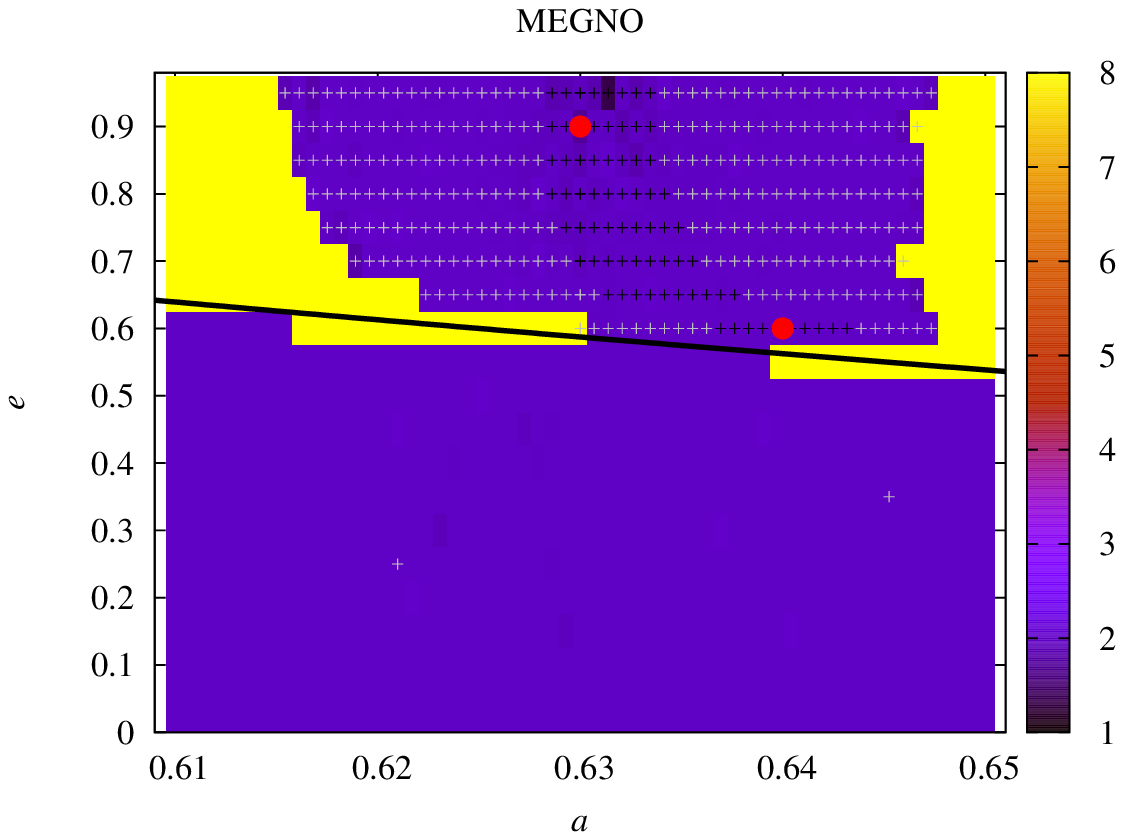}
% figure caption is below the figure
\caption{Stability map showing  regular (blue) and chaotic (yellow) regions for retrograde 2/1 resonance  with  $\phi_{21}^{\star}=0$ (top) or $\phi_{21}^{\star}=180^{\circ}$ (low) when $\mu=0.001$. The symbol $+$ marks initial conditions such that $\phi_{21}^{\star}$ librates around 0 (top) or $180^\circ$ (low) with amplitude less than $50^{\circ}$ (black $+$), less than $300^{\circ}$ (gray $+$). The apocentric collision line is shown in black. The initial conditions corresponding to orbits in Fig.~\ref{xy21} (left) are marked with red circles on the top panel, while
 those corresponding to orbits in Fig.~\ref{xy21} (right) are marked with red circles on the low panel. The 2 gray $+$ symbols below the apocentric collision line (low panel) correspond to non-resonant orbits that are wrongly associated with large amplitude libration due to output sampling.}
\label{megnoR21}       % Give a unique label
\end{figure*}

 The test particle's initial conditions for the 2/1 retrograde resonance are: inclination $I=180^\circ$, semi-major axis in $(0.63-0.3\,R_H,0.63+0.3\,R_H)$ varying at steps $0.01\,R_H$, eccentricity in $(0,1)$ varying at steps $0.05$, $\Omega=\omega=0$, $\phi_{21}^{\star}=\lambda^\star-2\,\lambda^\prime-3\,\varpi^\star=0,180^\circ$, with $\lambda^\star=\lambda-2\,\Omega$, $\varpi^\star=\omega-\Omega$.  In Fig.~\ref{megnoR21} we show the MEGNO maps for configurations with $\phi_{21}^{\star}=0$ (top panel) and $\phi_{21}^{\star}=180^\circ$ (low panel). The initial $(a,e)$ corresponding to the orbits in Fig.~\ref{xy21} (left) are marked in Fig.~\ref{megnoR21} (top), while
 those corresponding to the orbits in Fig.~\ref{xy21} (right) are marked in Fig.~\ref{megnoR21} (low).

The 2/1 retrograde resonant mode with $\phi_{21}^\star$ librating around 0  has close approaches with the planet between pericentre and apocentre (Fig.~\ref{xy21}, left) while the one with   $\phi_{21}^\star$ librating around $180^\circ$ has closest approaches with the planet at apocentre (Fig.~\ref{xy21}, right). Hence the apocentric collision line (black)  limits the  border of the resonance in Fig.~\ref{megnoR21} (low panel). Libration in the resonance with centre $\phi_{21}^\star=180^\circ$ is only possible above this collision line and the border is chaotic (yellow)  except for a gap between $0.63<a<0.64$ connecting the regular resonant and non-resonant regions. 
  
% For two-column wide figures use

\subsection{The 1/2 retrograde resonance}

\begin{figure*}
\centering
  \includegraphics[width=0.95\textwidth]{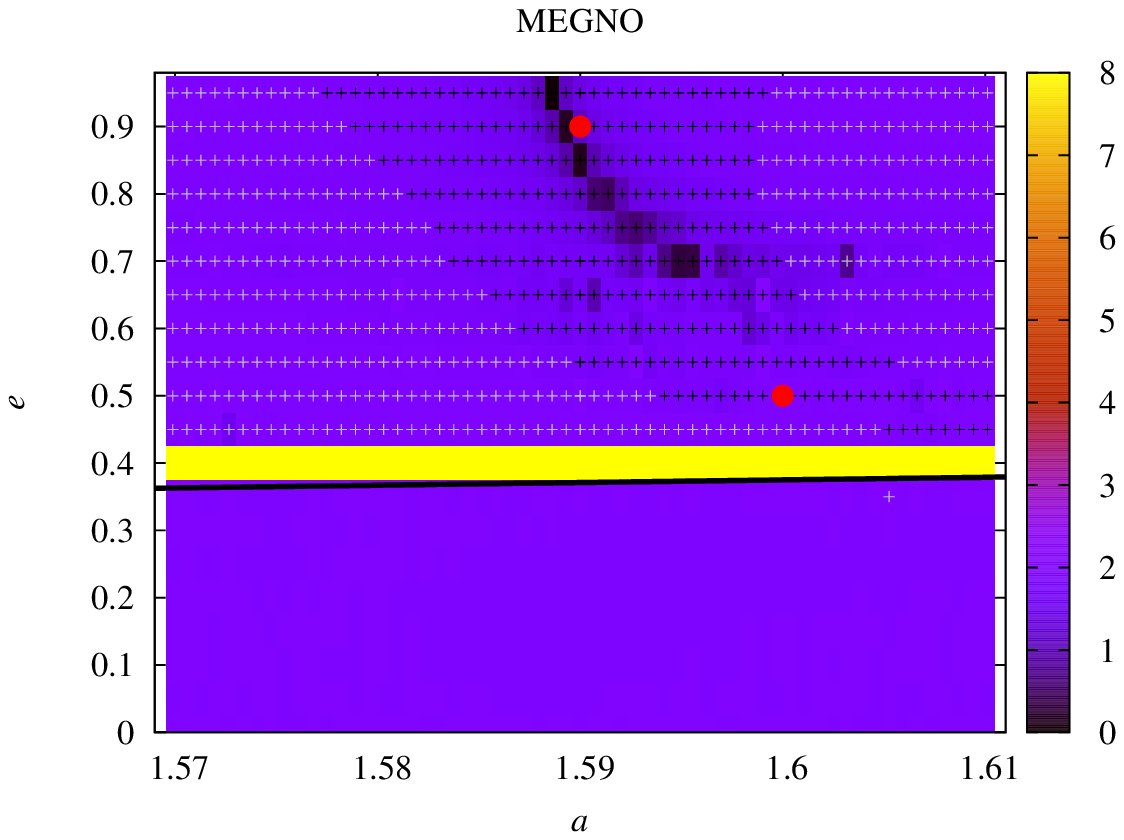}
  \includegraphics[width=0.95\textwidth]{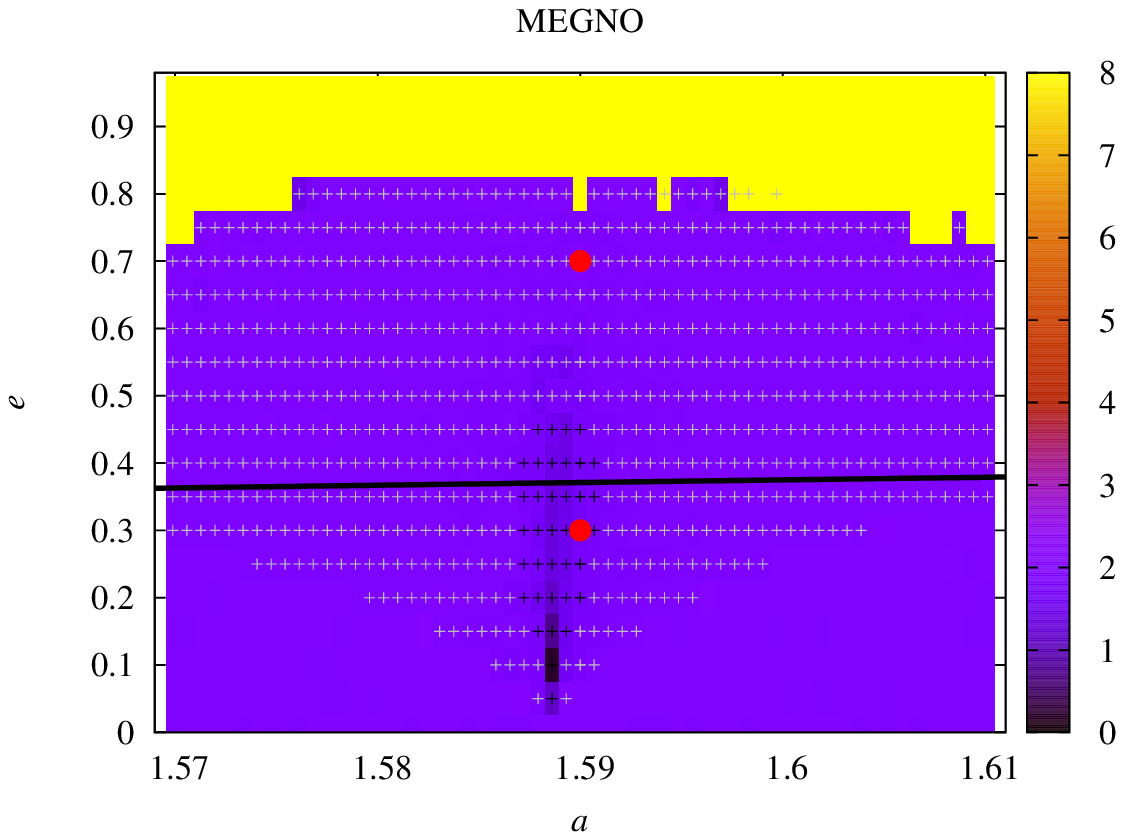}
% figure caption is below the figure
\caption{Stability map showing  regular (blue) and chaotic (yellow) regions for retrograde 1/2 resonance  with  $\phi_{12}^{\star}=0$ (top) or  $\phi_{12}^{\star}=180^{\circ}$ (low) when $\mu=0.001$. The symbol $+$ marks initial conditions such that $\phi_{12}^{\star}$ librates around 0 (top) or $180^\circ$ (low) with amplitude less than $50^{\circ}$ (black $+$), less than $300^{\circ}$ (gray $+$). The pericentric collision line is shown in black. The initial conditions corresponding to orbits in Fig.~\ref{xy12} (left) are marked with red circles on the top panel, while
 those corresponding to orbits in Fig.~\ref{xy12} (right) are marked with red circles on the low panel. 
 The gray $+$ symbol just below the pericentric collision line (top panel) correspond to a non-resonant orbit that is wrongly associated with large amplitude libration due to output sampling.}
\label{megnoR12}       % Give a unique label
\end{figure*}

 The test particle's initial conditions for the 1/2 retrograde resonance are: inclination $I=180^\circ$, semi-major axis in $(1.59-0.3\,R_H,1.59+0.3\,R_H)$ varying at steps $0.01\,R_H$, eccentricity in $(0,1)$ varying at steps $0.05$, $\Omega=\omega=0$, $\phi_{12}^{\star}=2\,\lambda^\star-\lambda^\prime-3\,\varpi^\star=0,180^\circ$, with $\lambda^\star=\lambda-2\,\Omega$, $\varpi^\star=\omega-\Omega$.  In Fig.~\ref{megnoR12} we show the MEGNO maps for configurations with $\phi_{12}^{\star}=0$ (top panel) and $\phi_{12}^{\star}=180^\circ$ (low panel).  The initial $(a,e)$ corresponding to the orbits in Fig.~\ref{xy12} (left) are marked in Fig.~\ref{megnoR12} (top), while
 those corresponding to the orbits in Fig.~\ref{xy12} (right) are marked in Fig.~\ref{megnoR12} (low).

The 1/2 retrograde resonant mode with $\phi_{12}^\star$ librating around 0  has close approaches with the planet at pericentre (Fig.~\ref{xy12}, left) while the one with  $\phi_{12}^\star$ librating around $180^\circ$ has closest approaches with  planet between apocentre and pericentre (Fig.~\ref{xy12}, right). Hence the pericentric collision line (black) limits the border of the resonance in Fig.~\ref{megnoR12} (top panel). Libration in the resonance with centre $\phi_{12}^\star=0$ is only possible above this collision line which coincides with a chaotic (yellow) strip separating the regular resonant and non-resonant regions.

% For two-column wide figures use

 \section{Discussion}

In this article we extended our work on the stability of retrograde configurations and the nature of retrograde resonance in the framework of the planar circular restricted three-body problem (CR3BP) consisting of a star, planet and test particle \cite{Morais&Giuppone2012MNRAS,Morais&Namouni2013CMDA}.
 
We obtained the stability boundary for retrograde configurations in the low mass (planetary) regime and showed that it is  located within the planet's Hill's radius and that there is an asymmetry between the inner and outer boundaries. In contrast, the stability boundary for prograde configurations is located at several planet's Hill's radius and is approximated by Wisdom's criterion of first order orbital resonances' overlap. 
These results help understand the numerical simulations in \cite{Namouni&Morais2015MNRAS} that showed that smooth migration capture in the 1/1 (coorbital) resonance occurs with probability 1 for planar retrograde configurations but  with probability 0  for planar prograde configurations. We leave further investigation of this capture mechanism for a future publication

We obtained stability maps in a grid of semi-major axis versus eccentricity for the 2/1 and 1/2 retrograde resonances and showed the regions of libration of the critical arguments.  The resonance borders are delimited by the collision separatrix  with the planet whose location depends on the geometry of the resonance configuration and does not necessarily occur at pericentre (apocenter) for outer (inner) resonances. In the configurations where collision is possible at apocentre or pericentre our results agree with a study of prograde 3rd and 4th order resonances in the planar CR3BP by \cite{Erdi2012CMDA}.

Stability studies of individual retrograde resonances are important in order to understand the origin of the Solar System small bodies in retrograde resonance with Jupiter and Saturn identified by \cite{Morais&Namouni2013MNRASL}. Retrograde resonances may also exist in extrasolar systems \cite{Gayon&Bois2008,Gayon&Bois2009MNRAS,Gozdziewski2013MNRAS} although no cases have yet been confirmed due to the difficulty in measuring relative inclinations.

\begin{acknowledgements}
We would like to acknowledge the assistance of Nelson Callegari Jr.~with computational resources.
\end{acknowledgements}

% BibTeX users please use one of
%\bibliographystyle{spbasic}      % basic style, author-year citations
\bibliographystyle{spmpsci}      % mathematics and physical sciences

\bibliography{retrogradestability}  % name your BibTeX data base

\end{document}